\documentclass[superscriptaddress,twocolumn,aps,pra]{revtex4-1}
\usepackage{graphicx}
\usepackage{dcolumn}
\usepackage{bm}
\usepackage[utf8]{luainputenc} \usepackage{amsmath}
\usepackage{amssymb} \usepackage{graphicx}
\usepackage{caption} \usepackage{float}

\makeindex

\begin{document}

\title{On-chip generation and control of the vortex beam}

\author{Aiping Liu}
\address{College of Telecommunications and Information
Engineering, Nanjing University of Posts and
Telecommunications, Nanjing, Jiangsu 210000, China. Key Lab of Broadband Wireless Communication and Sensor Network Technology, Nanjing University of Posts and Telecommunications, Ministry of Education, Nanjing 210003, China}

\author{Chang-Ling Zou}
\email{clzou321@ustc.edu.cn}
\address{Key Laboratory of Quantum Information, University
of Science and Technology of China, CAS, Hefei, Anhui
230026, China. Synergetic Innovation Center of Quantum
Information $\&$ Quantum Physics, University of Science and
Technology of China, Hefei, Anhui 230026, China}
\address{Department of Electric Engineering, Yale University, New Haven, CT 06511, USA}

\author{Xifeng Ren}
\email{renxf@ustc.edu.cn}
\address{Key Laboratory of Quantum Information, University
of Science and Technology of China, CAS, Hefei, Anhui
230026, China. Synergetic Innovation Center of Quantum
Information $\&$ Quantum Physics, University of Science and
Technology of China, Hefei, Anhui 230026, China}

\author{Qin Wang}
\address{College of Telecommunications and Information
Engineering, Nanjing University of Posts and
Telecommunications, Nanjing, Jiangsu 210000, China. Key Lab of Broadband Wireless Communication and Sensor Network Technology, Nanjing University of Posts and Telecommunications, Ministry of Education, Nanjing 210003, China}

\author{Guang-Can Guo}
\address{Key Laboratory of Quantum Information, University
of Science and Technology of China, CAS, Hefei, Anhui
230026, China. Synergetic Innovation Center of Quantum
Information $\&$ Quantum Physics, University of Science and
Technology of China, Hefei, Anhui 230026, China}

\date{\today}

\begin{abstract} A new method to generate and control the amplitude and phase distributions of a optical vortex beam is proposed. By introducing a holographic grating on top of the dielectric waveguide, the free space vortex beam and the in-plane guiding wave can be converted to each other. This microscale holographic grating is very robust against the variation of geometry parameters. The designed vortex beam generator can produce the target beam with a fidelity up to $0.93$, and the working bandwidth is about $175$ nm with the fidelity larger than $0.80$. In addition, a multiple generator composed of two holographic gratings on two parallel waveguides are studied, which can perform an effective and flexible modulation on the vortex beam by controlling the phase of the input light. Our work opens a new avenue towards the integrated OAM devices with multiple degrees of optical freedom, which can be used for optical tweezers, micronano imaging, information processing, and so on.
\end{abstract}

\maketitle

Optical vortex beams with phase singularities were firstly
proposed by Nye and Berry in 1974 \cite{1Nye74} and proved
to carry Orbital Angular Momentum (OAM) by Allen \textit{et al.} in
1992 \cite{2Allen92}. OAM is an intrinsic character of photon
and allows the optical beam have a helical phase front and
a phase singularity. Beams carrying OAM have the azimuthal
angular dependence of $exp(il\theta)$, where $l$ is the
azimuthal index and $\theta$ is the azimuthal angle. Since
different OAMs are orthogonal to each other, thus OAMs of
photons provide an alternative degree of freedom to encode
information, for both classical and quantum mechanics. In principle,
$l$ is an unbounded number, allowing the single photon to carry
high dimensional information. Therefore, OAM has drawn interests
for the potential applications in communication
\cite{3Gibson04,DAmbrosio,yuan15}, optical microscopy
\cite{5Furhapter05}, remote sensing \cite{6Tamburini11} and
quantum information \cite{7Mair01,Ren,Ren06,ShiBS15,LiChF15}. In particular, the
OAM tweezers can be used as optical spanner to trap and
rotate nanomechanics, atom ensembles and nanobio-molecules
\cite{Grier,Paterson,4Curtis03}.

There are different approaches to prepare and detect the vortex
beams, such as diffraction gratings
\cite{Torner05,Monroe15}, spiral phase mirror
\cite{Oemrawsingh04} and spatial liquid phase modulator
\cite{Savage07,Chenlixiang14}. However, these methods are
free space components, which occupy $\mathrm{cm}^3$ spaces,
thus limiting the further scalability. Alternative approaches
based on the whispering gallery modes in microsphere
\cite{Ilchenko} and microring \cite{10Cai12} cavities and
plasmonic structures \cite{Liu12,Liu13} are proposed and
demonstrated. Light traveling in the cavities can be converted
to free space beam with non-zero OAM, which is promising for
the integration on the photonic chips, but only works for specific
wavelengths with a very narrow bandwidth.

Here, we propose a scheme of generating and manipulating vortex
beams by integrated waveguides. The free space vortex beams are
coupled with the guiding modes in the dielectric waveguides through
diffraction elements, which is designed on the top surface of the waveguides by the holographic method. For single
waveguide, high quality vortex beam with $l=1$ are generated,
with the waist size of 0.5\,$\mu $m. The performance of the
integrated devices is very robust against the variation of width and
length of diffraction gratings. Remarkably, the working
wavelength is broadband with bandwidth of about 175\,nm for
the 670\,nm-holographic grating.
The most attractive advantage of the proposed vortex beam generator is the scalability, which permits multiple generator composed of a two-dimensional array of gratings fabricated on a single chip. The multiple generator allows the precise and real-time control of the beam. As an example, we demonstrate the manipulation of the beam shape and position
by just controlling  wavelength and phase of the input lasers to two waveguides.

\begin{figure}[htbp]
\centering
\includegraphics[width=\linewidth]{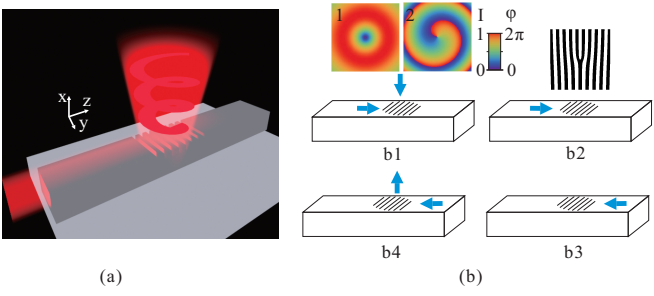}
\caption{(color online) (a) The schematic illustration of the proposed vortex beam generator on the integrated waveguide.
(b1)-(b4) The principle of the holographic grating on waveguide.}
\label{fig:1}
\end{figure}

The schematic of the proposed vortex beam generator is shown in Fig.\,1(a), in which the incident light from one port of the waveguide (in plane) will be converted to free space (in vertical direction) vortex beam by the waveguide surface holographic grating (WGSHG). The WGSHG is essential for our device, since it determines the qualities of the generated beams. The basic principle of the holographic grating is explained in Figs. 1(b1)-(b4). At first step, the time-reversal of the target vortex beam is incident vertically to waveguide, and at the same time, there is also a guiding wave propagating from left to right in the waveguide (Fig.\,1(b1)). The interference of the two wave fields leads to string with dark and bright patterns on the surface of the waveguide. The WGSHG is generated by dividing the interference region into many pixels ($50\,\mathrm{nm}\times50\,\mathrm{nm}$) and setting a gray value of  $G(x,y)$ for each pixel \cite{11Chen12}. Here, we use the simple binary function according to the phase difference $\Delta\theta$ between the target vortex beam and the guiding wave, i.e.  $G(x,y)=0$ if $|\Delta\theta|<0.5\pi$, or else $G(x,y)=1$. The inset of Fig. 1(b2) gives the binary gray image of the holographic grating. Due to the principle of holography, when the WGSHG is fabricated, a conjugate guiding wave propagating along the waveguide (the direction of the blue arrow is from right to left in Fig.\,1(b3)) will produce the target vortex beam. Comparing the intensity and phase distributions of the input target beam and the generated beam, the reversal of the phase helicity indicates the time-reversal relation of two beams as a manifest of holographic principle. Similarly, if a vortex beam is incident from the bottom of the waveguide, it will be converted into guiding wave in the waveguide traveling from right to left. Thus, this would be potential for vortex beam detection.

In the following, we verify this idea by numerical simulations. The waveguide with a width of $b$ is  made of $\mathrm{Si_{3}N_{4}}$ on a silica substrate. The refractive index of $\mathrm{Si_{3}N_{4}}$ is taken as 2.0063 at the wavelength of $\lambda_{0}=670$\,nm. The electric field in the waveguide is in the form as $E_{x,y,z}=A(x,y)e^{-i\kappa_{z}z}$ and can be solved numerically, where $A(x,y)$ is the amplitude of the electric field in the $ xy $-cross section, $\kappa_{z}$ is the wave vector of guiding wave in the $ z $-direction. In current study, we focus on the fundamental TE mode, whose field distribution on the top surface of waveguide can be approximated by a Gaussian function for $d>\lambda$. The target vortex beam with OAM $l=1$ is shown in the insets of Fig.\,1(b4), containing a donut amplitude distribution and a helical phase front with a singularity in the center. The waist of the target vortex beam is posited on the up surface of the waveguide with a diameter of 500\,nm. The geometry of the WGSHG is a rectangle with a width of $b$ (same as waveguide) and a length of $d$.  In the experiment, the dielectric nanostructure of the WGSHG can be fabricated by etching the surface layer of the waveguide \cite{Taillaert02}, or by nano-imprinting method \cite{Chou96}. More intriguing way to do so is to introduce a metallic nanostructure \cite{Arango}, which forms a metasurface \cite{10Cai12,12Ni13,Montelongoa14}, allowing efficient light extraction and precisely control of the local scattering phase at subwavelength.

\begin{figure}[htbp]
\centering
\includegraphics[width=\linewidth]{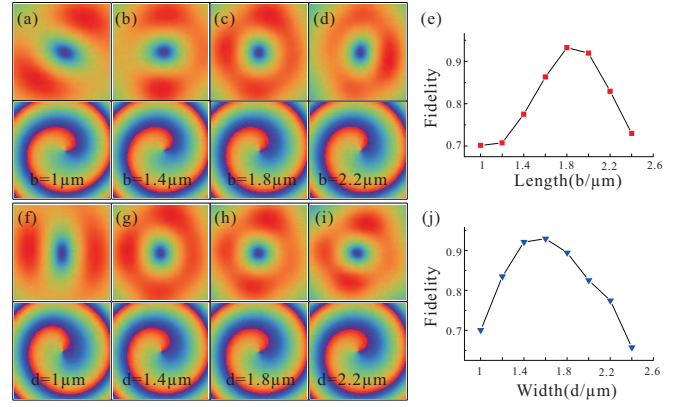}
\caption{(color online). The obtained vortex beams by
the holographic gratings with different sizes. The intensity
and phase distributions are shown in the up part and down
part of the figures, (a)-(d) corresponding to
the holographic gratings with a fixed length $ d=1.5\,\mu\mathrm{m}$ but
different widths $ b=1,\,1.4,\,1.8,\,2.2\,\mu\mathrm{m}$ and (f)-(i)
corresponding to the holographic gratings with a fixed width
$ b=1.8\,\mu\mathrm{m}$ but different lengths $ d=1,\,1.4,\,1.8,\,2.2\,\mu\mathrm{m}$, respectively. (e) and (j) are the fidelities of the
obtained vortex beams as the functions of width b and length d,
respectively. }
\label{fig:2}
\end{figure}

In order to obtain a vortex beam with high quality, the size effect of the holographic grating is studied at first. Figs. 2(a)-2(d) give the simulation results of the obtained beam for the holographic gratings with a fixed length $d=1.5\,\mu\mathrm{m}$ and different width $ b=1,\,1.4,\,1.8,\,2.2\,\mu\mathrm{m}$, respectively. Here and in the following of this paper, the field distribution of the vortex beam are on the plane with a distance of $ 10\lambda_{0} $ away from the top surface of the waveguide. The up and down parts of the figures are the intensity and phase
distributions, respectively. Similar to the target vortex beam, the obtained beam has a null amplitude on the center and a helical phase distribution, which indicates that the obtained beam carries OAM as the target vortex beam. When the width $b$ is changed, the phase distribution keeps the same helical distribution, and the amplitude distribution varies, which means that the fidelity of the obtained vortex beam is related to the width of the holographic grating. To qualify the quality of the generated vortex beam, the fidelity is introduced
\begin{equation}
\mathcal{F}=\frac{\lvert\int A_{o}^{*}(x,y,z)A_{t}(x,y,z)dxdydz\rvert^{2}}{\int\lvert A_{o}(x,y,z)\rvert^{2}dxdydz\int\lvert A_{t}(x,y,z)\rvert^{2}dxdydz},
\label{eq:2}
\end{equation}
where $A_{o}(x,y,z)$ and $A_{t}(x,y,z)$ are the amplitudes of the obtained and target vortex beams, respectively. Because of the finite number of imaging pixels, the difference exists between the target beam and the obtained beam, which makes the fidelity $\mathcal{F}<1 $. Fig.\,2(e) gives the fidelities of the vortex beams as the function of the width $b$, where the fidelities are above 0.7 for $b$ in the range of $[1,2.4]\,\mu$m. In fact, the fidelity increases with b to reach a highest one, and then decreases with b. The best fidelity
is 0.93 when $ b=1.8\,\mu$m, which is used as the width of the holographic grating in the following simulation. Figs 2(f)-2(j) studies the generation of the vortex beam depending on the grating length $d$ with the fixed width $ b=1.8\,\mu$m. From the field distributions for $d=1,\,1.4,\,1.8,\,2.2\,\mu$m, $d$ has little effect on the phase distribution, but obvious effect on the amplitude distribution. The fidelity shown in Fig.\,2(j) has an optimum $\mathcal{F}=0.93$ for $d=1.6\,\mu$m.

\begin{figure}[htbp]
\centering
\includegraphics[width=1\linewidth]{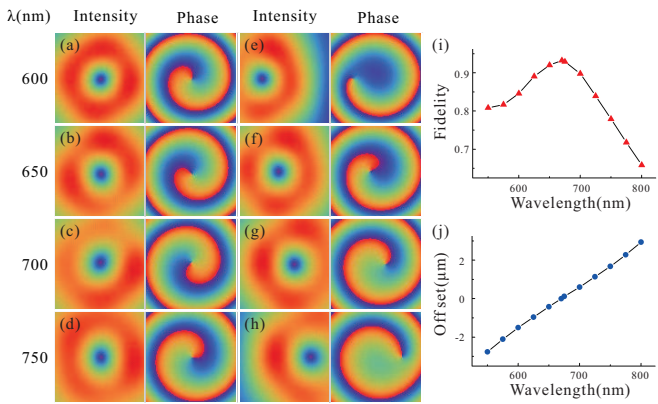}
\caption{(color online).The intensity and phase
distributions of the obtained vortex beams with different
wavelengths. (a)-(d) are the intensity (left) and phase
(right) distributions for the holographic gratings (or guiding waves) with $\lambda=600,\,650,\,700,\,750\,\mathrm{nm}$, respectively. (e)-(h) are the intensity (left)
and phase (right) distributions for the guiding waves with
$\lambda=600,\,650,\,700,\,750\,\mathrm{nm}$, respectively,
obtained from the holographic grating with $
\lambda=670\,\mathrm{nm} $. (i) and (j) are the fidelity and the off set as the functions of wavelength, respectively. }
\label{fig:3}
\end{figure}

Use the optimum parameters obtained from the above results, we can produce vertex beam with high quality by the on-chip waveguide. Further, we studied the generation of vertex beam with different wavelength and found that the WGSHG is fascinating, because it can work in the wavelength ranged among $ 450-900\,\mathrm{nm}$. Figs.\,3(a)-(d) give the intensity (left) and phase (right) distributions of the obtained vortex beams for the wavelengths $ \lambda=600,\,650,\,700$ and $750\,\mathrm{nm}$, with fidelities $\mathcal{F}=0.91,\,0.84,\,0.86,\,0.93$, respectively. Once the holographic grating is fabricated, the geometry of the holographic grating is determined, but it can still be applied to a broadband wavelength. Figs. 3(e)-(h) give the intensity (left) and phase (right) distributions of the obtained vortex beams for the holographic gratings with wavelength $\lambda=670\,\mathrm{nm}$, and the guiding waves with $\lambda=600,\,650,\,700, \,750\,\mathrm{nm}$, respectively. The direction of the obtained vortex beam is no longer vertical to
the waveguide, when the wavelenth of the guiding wave does not match with that of the holographic grating. The amplitude and phase distributions are off set from the wavelength-matched case. Figs. 3(i) and 3(j) give the fidelity and off set as the functions of the guiding wavelength, respectively. The fidelity decreases as the guidin wavelength is far away from that of the holographic grating. For the guiding wavelength among $550\,\mathrm{nm}-725\,\mathrm{nm} $, the fidelities are all above $0.80$, which indicates the holographic grating is a brandband grating. The off set varies with the guiding wavelenth linearly, where it is minus for shorter wavelength and positive for longer wavelength, respectively. According to this linear ralation, the position of the obtained vortex beam can be controlled by manipulating the guiding wavelength, which improves the flexibility of the vortex beam generator.

To control the obtained vortex beam more flexibly, a multiple generator, composed of two gratings on top of two parallelly arrayed waveguides, are designed to generate the vortex beam. Figure 4(a) gives the schematic of generating vortex beam by the multiple generator, with the illustration of the holographic gratings on the waveguides in the inset. The width of each grating is 1 $\mu$m, which is the same with that of each waveguide. The length of each grating is 2.5 $\mu$m. The gap between the two gratings (or the waveguides) is 0.5 $\mu$m, which makes the generated vortex beam have a high quality. There is a phase difference $\Delta\theta$ between the guiding waves in the waveguides A and B. Lights diffracted from different gratings will interfere with each other to form different intensity distributions according to $\Delta\theta$. So, the generated beam can be controlled by manipulating the phase difference between the guiding waves in the two waveguides. Figs. 4(b)-4(k) give the intensity and phase distributions of the vortex beams for various phase difference. When they are in-phase $\Delta\theta=0$, the obtained beam has a null amplitude point and a phase singulary point on the center, which is the typical characteristic of a vortex beam. When $\Delta\theta$ increases, the null amplitude point moves, and another null amplitude point appears gradually. For $\Delta\theta=\pi$ (Fig.\,4(d)), there are two clear null amplitude points. While $\Delta\theta$ increases further, one of them disappears and only the other one exists. The phase singularity performs the same phenomena when $\Delta\theta$ varies (Figs. 4(g)-(k)). Since the phase difference of the guiding waves in the two waveguides can be manipulated arbitrarily and precisely, the reproduced light can be controlled flexibly, which allows the realization of manipulating the vortex beam.

\begin{figure}[htbp]
\centering
\includegraphics[width=1\linewidth]{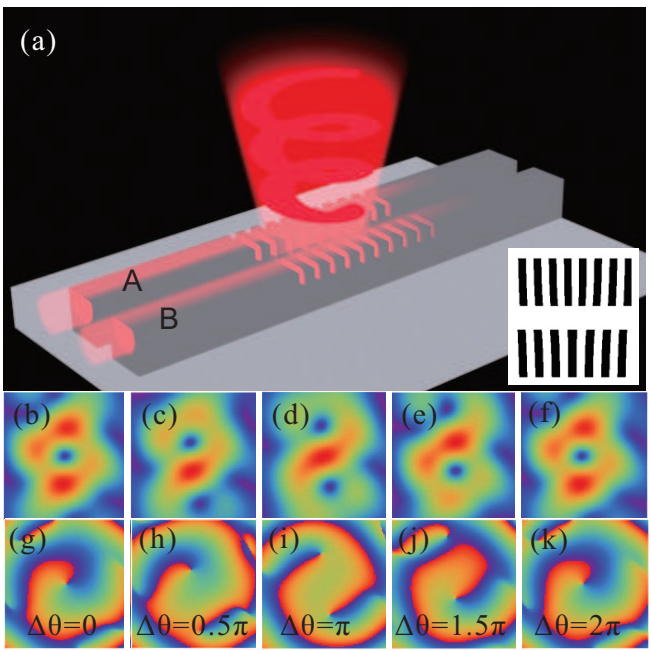}
\caption{(color online). The schematic of generating and modulating vortex beam with a multiple generator. (a) The genetation of vortex beam by the multiple generator composed of two gratings on top of two parallelly arrayed waveguides A and B. The inset is the holographic gratings on the waveguides. The intensity (b)-(f) and phase (g)-(k)
distributions of the obtained vortex beams for the guiding waves
with phase differences $\Delta\theta=0,\,0.5\pi,\,\pi,\,1.5\pi,\,
2\pi$, respectively.} \label{fig:4}
\end{figure}

In conclusion, the generation and modulation of the vortex beam
are realized by the holographic grating on the waveguide. The
quality of the obtained vortex beam is improved by optimizing
the size of the holographic grating, and the fidelity of
the obtained vortex beam is up to 0.93. The holographic grating
is proved to be broadband with a bandwidth of 175 nm for the
670 nm-holographic grating. The direction of the generated
vortex beam can be controlled by the guiding wavelength
linearly. Further, the modulation of the obtained vortex beam
can be realized by the multiple generator, which shows that
both amplitude and phase of the obtained vortex beam can be modulated effectively through the phase difference between the guiding waves
in the two waveguides. The generation and modulation of vortex
beam by the holographic grating on the waveguide expands a
route to provide light source with OAM on a compact chip, which will
play a key role in the integrated optics.

\section*{Acknowledgments}

National Basic Research Priorities Program of China (2011CBA00200, 2011CB921200);
The “Strategic Priority Research Program(B)” of the Chinese Academy of Sciences (XDB01030200);
The National Natural Science Foundation of China (11504183, 11374289, 11274178, 61475197);
The Fundamental Research Funds for the Central Universities (WK2470000012);
The Natural Science Foundation of the Jiangsu Higher Education Institutions (15KJA120002, BK20150039);
The Scientic Research Foundation of Nanjing University of Posts and Telecommunications (NY214142);



\bigskip{}

\noindent 

\end{document}